\documentclass[%
 reprint,
 amsmath,amssymb,
 aps,
]{revtex4-1}

\usepackage{graphicx}% Include figure files
\usepackage{dcolumn}% Align table columns on decimal point
\usepackage{bm}% bold math
\usepackage{hyperref}% add hypertext capabilities
\begin{document}

\preprint{APS/123-QED}

\title{Antineutrino flux from the Laguna Verde Nuclear Power Plant}% Force line breaks with \\

\author{Marisol Ch\'avez-Estrada}
\email{fismarisol@gmail.com}
\author{Alexis A. Aguilar-Arevalo}
 
\affiliation{
 Instituto de Ciencias Nucleares. Universidad Nacional Aut\'onoma de M\'exico \\
 Circuito Exterior, Ciudad Universitaria, Apartado postal 70-543, 04510  D.F., M\'exico 
}

\begin{abstract}
\noindent We present a calculation of the antineutrino flux produced by the reactors at the Laguna Verde Nuclear Power Plant in M\'exico, based on the antineutrino spectra produced in the decay chains of the fission fragments of the main isotopes in the reactor core, and their fission rates, that have been calculated using the DRAGON simulation code. We also present an estimate of the number of expected events in a detector made of plastic scintillator with a mass of 1 ton, at 100 m from the reactor cores.\\

\textit{Keywords:} Reactor antineutrinos, Laguna Verde, DRAGON simulation, nuclear safeguards.\\%de 1 a 5 keywords
\\
%Se presenta una estimaci\'on del flujo de antineutrinos producido por los reactores de la Central Nucleoel\'ectrica de Laguna Verde en M\'exico a partir de los espectros de energ\'ia de los antineutrinos producidos en las cadenas de decaimiento de los fragmentos de fisi\'on de los is\'otopos principales en el n\'ucleo del reactor, y de las tasas de fisi\'on de estos, calculadas con el c\'odigo de simulaci\'on DRAGON. Se presenta una estimaci\'on del n\'umero de eventos esperados en un detector de pl\'astico centellador de 1 ton, a 100 m de los n\'ucleos de los reactores.\\

%\textit{Descriptores:} Neutrinos, Laguna Verde, DRAGON, salvaguardias nucleares, decaimiento beta inverso.\\

%PACS: 13.15.+g; 14.60.Lm; 28.41.-i. %1 a 3 codigos PACS http://www.aip.org/pacs  Physics and Astronomy Classification Scheme.

\end{abstract}

\maketitle
\section{\label{sec:Introduccion}Introduction}
%Since the discovery of antineutrinos in the 1950s by Cowan and Reines \cite{Cowan_Reines}, using the nuclear reactor at Savannah River in South Carolina, there have been many experiments around the world that have characterized the properties of these particles, using the fact that fission reactors are an intense source of electron antineutrinos. 

In recent years, there has been significant progress in understanding the properties of neutrinos, most notably, since 2012 the results of the reactor neutrino experiments Daya Bay~\cite{DayaBay}, Reno and Double Chooz~\cite{DoubleCHOOZ} have determined that the mixing angle $\theta_{13}$ is nonzero with a high level of significance. The current degree of development of neutrino detection technology is close to making a reality to non-intrusively monitor the operational status, power level and fissile content of a nuclear reactor in real time using detectors placed at distances of a few tens of meters.

Monitoring nuclear reactors through its antineutrino flux is a complimentary and promising new tool for supervising the operations of nuclear plants, which are bound to operate according to protocols established by the International Atomic Energy Agency (IAEA). This agency is responsible for ensuring that nuclear reactors world wide operate legally, preventing the diversion of fissile material to activities which could lead to the manufacture of weapons.

Many research groups worldwide (~\cite{SONGS}-\cite{DANSS}) have studied this application aiming at adding antineutrino detection to the techniques used to implement reactor safeguards. In a medium term plan (5 to 8 years) IAEA~\cite{IAEA_workshop} has proposed that antineutrino detectors for safeguard applications should be able to provide information on the thermal power, fissile content and operational status of reactors while deployed aboveground in a compact volume to minimize its intrusiveness with the nuclear power plant operations. The SONGS1 experiment~\cite{SONGS} using a Gd loaded liquid scintillator detector, deployed at 25 m from the reactor core of the San Onofre Nuclear Generation Station, successfully proved that the reactor ON/OFF cycle and the thermal power stability of a reactor can be measured using a compact and simple detector. PANDA [4], lead by Tokyo University, used a segmented plastic scintillator detector with modules wrapped with Gd coated sheets in between. This type of detector makes it possible to use information of the event topology to tag antineutrino events and to discriminate them from backgrounds such as fast neutrons. Furthermore, it can be transported and operated easily inside a compact vehicle, and has the advantage that unlike the liquid scintillators, is non-flammable.

M\'exico has two nuclear fission reactors at the Laguna Verde Nuclear Power Plant (LVNPP), which operates commercially since 1990 (Unit 1) and 1995 (Unit 2). This is the only nuclear power plant in the country and generates about 5\% of its total electric power production. In this work we calculate the antineutrino flux produced by these reactors and give some estimates of the number of events expected in a generic plastic scintillator detector located at a distance of 100 m from the reactor cores. Knowledge of the antineutrino flux produced by a particular reactor is an important step towards its monitoring using this method. 

While abundant literature exists about the calculation of neutrino fluxes from pressurized water reactors (PWR)~\cite{literature}, much fewer information is available about neutrino production at boiling water reactors (BWR) (See for example~\cite{Nakajima}). These two reactor types operate under the same basic physical principles regarding nuclear fission, but differences in their operational procedures may effect differently the neutrino flux they produce, as well as its evolution along the reactor fuel cycle.

This paper is organized as follows: in section \ref{sec:CNLV} main technical data on the LVNPP and its reactors are detailed. In section \ref{sec:DRAGON} a description about the production of antineutrinos from nuclear reactors is given, and the calculation of antineutrino flux is presented. In Section \ref{sec:Flujo} parameters and characteristics of the DRAGON simulation code used to calculate fission rates of four major fissile isotopes for this paper are described. Section \ref{sec:eventos} shows the event rates or interactions that are expected to be observed with a 1 ton plastic scintillator detector placed at 100 meters from the LVNPP reactors. Finally, section\ref{sec:Conclusiones} presents the conclusions.

\section{\label{sec:CNLV}The Laguna Verde Nuclear Power Plant (LVNPP)}

The LVNPP is located in the municipality of Alto Lucero, Veracruz, M\'exico. This plant has two twin units, each equipped with a BWR-5 second generation reactor with a Mark II containment design, supplied by General Electric with a capacity of 2027 MWth, and net electrical output of 805 MWe. Both reactors operate with enriched uranium as fuel, and demineralized water as moderator and coolant. It is the only nuclear power plant in the country and generates about 5\% of the total electric power production in M\'exico~\cite{reporteIAEA}.

\begin{figure}[t] 
\centering    
\includegraphics[width=0.4\textwidth] {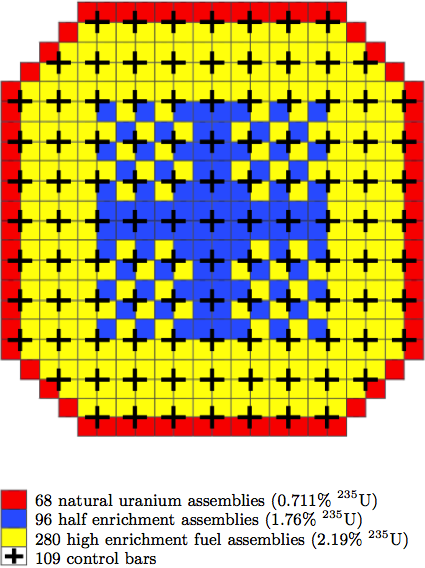}
\caption{\small{Schematic of a cross section of one of the LVNPP reactor cores \cite{folletoCNLV}. The "red" assemblies (0.711\%) are located on the perimeter of the reactor core, while the "blue" assemblies (1.76\%) are arranged in the central area. There are 109 control bars represented as crosses in the figure.}}
\label{fig:carga_inicial_3colores}
\end{figure}

Each nuclear reactor core can be approximately represented by a cylinder $\sim$4 m in height and 4 m in diameter containing 444 fuel assemblies and 109 control bars arranged as shown in Fig. \ref{fig:carga_inicial_3colores}. A fuel assembly is a square prism with sides of $\sim$13 cm and 4 m in height, containing an array of fuel rods (configurations of 8x8, 10x10 or 12x12 have been either used or considered). Within a given fuel assembly some of the rods contain uranium oxide (UO$_2$) with various levels of enrichment, others contain a mixture of UO 2 with gadolinium oxide (Gd$_2$O$_3$) at varying concentrations, and some of them are hollow in order to allow the flow of the refrigerant through them. Combining rods with different levels of enrichment and Gd$_2$O$_3$ concentrations permits to adjust the properties of the medium for the production, absorption and diffusion of neutrons differently at various location across the core.

\section{\label{sec:DRAGON}DRAGON simulation of the reactor core}

We used the numerical simulation code DRAGON~\cite{Pagina_DRAGON} with the modifications described in~\cite{modifica_DRAGON} to implement a simulation for the LVNPP reactors and extract the time evolution of the fission rates (number of fissions per second) of each of the main fisile isotopes in the core. Several research groups worldwide have implemented this simulation for PWR reactors, aiming at the same objective~\cite{fiss_frac_DRAGON}. The code solves the neutron transport equations of individual fuel assemblies, whose detailed composition is given as an input. The overall behaviour of neutrons across the core can be calculated by summing over the various types of assemblies that compose a given core configuration.

As described in section \ref{sec:CNLV}, a geometry with 444 fuel assemblies with 3 different types of $^{235}$U enrichment was considered. The initial fuel load that was assumed in this work is that shown in Fig. \ref{fig:carga_inicial_3colores}. Note that in this model no assembly contains plutonium at the beginning of the operation cycle.

\begin{figure}[b] 
\centering    
\includegraphics[width=0.48\textwidth] {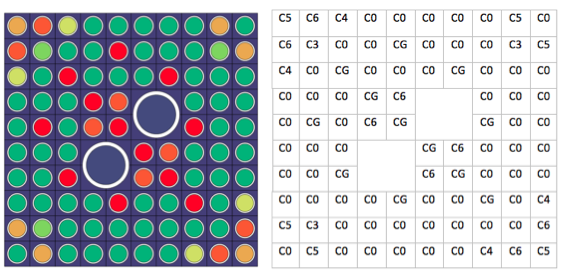}
\caption{\small{Arrangement of fuel rods for blue assembly (half enrichment). Left: DRAGON code output. Right: Bar type (see Table \ref{tab:ensamble_azul_tabla}). Notice that initially there are no fuel bars with Pu.}}
\label{fig:ensambleycelda}
\end{figure}

\begin{figure}[t] 
\centering    
\includegraphics[width=0.25\textwidth] {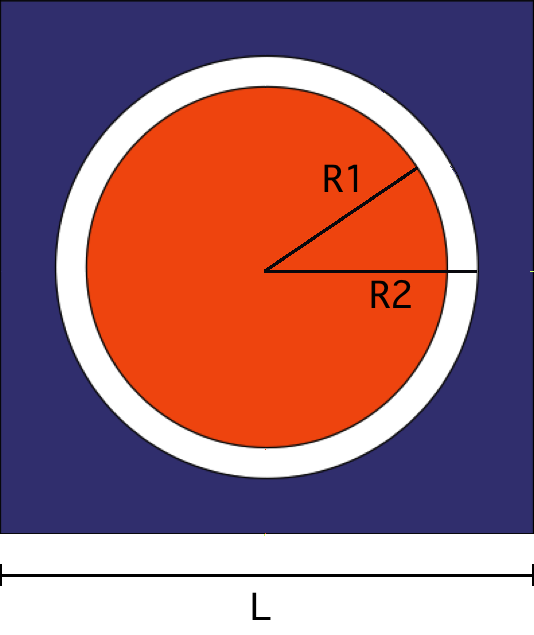}
\caption{\small{Cross section of a single unit cell. A fuel bar (in red) with a zircaloy cladding (white) immersed in demineralized water as moderator. A complete assembly is formed by 92 of these cells. The values for R1, R2 and L are described in Table \ref{tab:parametros_simulacion}}}
\label{fig:celda_unitaria}
\end{figure}

In Fig. \ref{fig:ensambleycelda}, we illustrate a single fuel assembly (in this case with a 10x10 rod configuration) corresponding to one with average enrichment of 1.76\% (blue assembly). The dimensions of a single unit cell, (the place occupied by a single fuel rod) are shown in Fig. \ref{fig:celda_unitaria}. The three types of assembly (red, yellow and blue) share the same geometry, however, the composition the fuel rods within a given assembly is, in general, different. The number of assemblies of each type, $N_A$ is: $N_{red} =68, N_{yellow}=280$ and $N_{blue}=96$, totaling 444.

We considered eight different fuel rod compositions, labeled $CX$, with $X = 0, 1, 2, 3, 4, 5, 6, G$, and used them to construct the various assemblies with the desired $^{235}$U enrichment and Gadolinium concentration. Only fuel rods of type G contain Gadolinium. Table \ref{tab:barras_por_ensamble} shows the number of fuel rods of each type that are used to build each of the three assemblies. As an example, Table \ref{tab:ensamble_azul_tabla} shows how we constructed the blue assembly.

\begin{table}[h]
	\begin{center} 
		\begin{tabular}{|c|c|c|c|c|c|c|c|c|c|} 
		\hline 
		Assembly & C0  & C1 & C2 & C3 & C4 & C5 & C6 & CG \\ 
		
		\hline
		Red & 92 & 0 & 0 & 0 & 0 & 0 & 0 & 0 \\
		\hline
		Blue & 59 & 0 & 0 & 3 & 4 & 6 & 8 &12\\
		\hline
		Yellow & 34 & 9 & 19 & 0 & 0 & 0 &0 &30\\
		\hline
				
		\end{tabular}
		\caption{\small{Number of fuel bars of type CX for each assembly. There is a total of 92 bars for each assembly.}}
		\label{tab:barras_por_ensamble}
	\end{center}
\end{table}

Table \ref{tab:parametros_simulacion} lists the values used for the various parameters required by the simulation. Tests showed a small dependence of the fission rates on variations of the coolant, moderator and fuel temperatures, while maintaining the thermal power constant.

A simulation was run for each of the three types of assembly (red, blue or yellow in Fig. \ref{fig:carga_inicial_3colores}), providing the fission rates $f_i^{(CX;A)}$ for each type of fuel rod, and for each one of the fissile isotopes $i=1,2,3,4$  ($^{235}$U, $^{238}$U, $^{239}$Pu and $^{241}$Pu, respectively).

To obtain the fission rates for a given assembly, $f^A_i$, we multiply the fission rate for each type of fuel rod times the number of rods of its type in the assembly, $N_{rods}$, and sum over all fuel rod types in the assembly, i.e:

\begin{equation}
f_i^A = \sum_{CX\in A} N_{rods} (CX;A) f_i^{(CX;A)}
\end{equation}

\begin{figure}[h] 
\centering    
\includegraphics[width=0.5\textwidth] {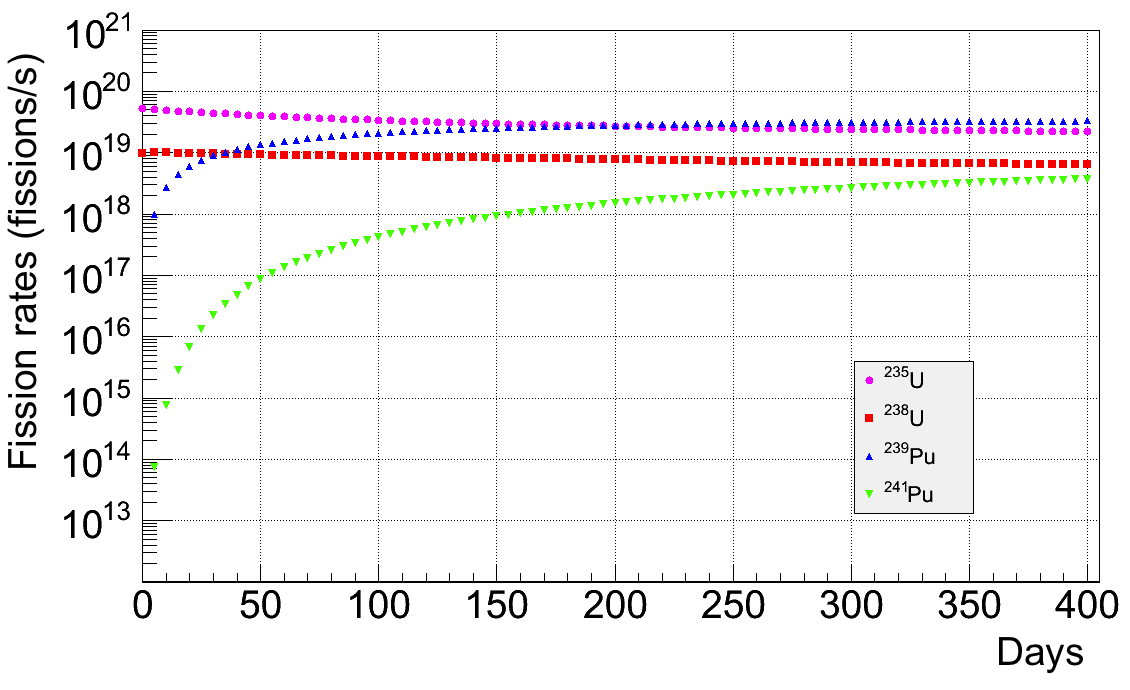}
\caption{\small{Fission rates (normalized) obtained with DRAGON for the 4 main fissile isotopes in the 444 fuel assemblies that make up the core of a BWR-5 of 2.027 GWth like the LVNPP reactor.}}
\label{fig:fissionrates}
\end{figure}

Finally, to calculate the fission rates of isotope $i$ in the complete reactor core, $f_i$ (Fig. \ref{fig:fissionrates}), we multiply the number of assemblies of each type $A$, times the corresponding
fission rates for that assembly:

\begin{equation}
f_i=\sum_{A} N_Af_i^A
\end{equation}

\begin{table}[t]
	\begin{center} 
		\begin{tabular}{|c|c|c|c|c|c|c|} 
		\hline 
		Rod type & \# of rods & $^{234}$U & $^{235}$U & $^{238}$U & O$_{16}$ & Gd \\ 
		   & & (\%)&(\%) &(\%) &(\%) &(\%)\\
		\hline
		C0 & 59 & 0.005 & 0.627 & 87.517 & 11.852 & -\\
		\hline
		C3 & 3 & 0.029 & 2.468 & 85.651 & 11.852 & -\\
		\hline
		C4 & 4 & 0.029 & 2.821 & 85.298 & 11.852 & -\\
		\hline
		C5 & 6 & 0.029 & 3.173 & 84.945 & 11.852 & - \\
		\hline
		C6 & 8 & 0.029 & 3.482 & 84.637 & 11.852 & - \\
		\hline
		CG & 12 & 0.028 & 3.142 & 79.691 & 11.933 & 5.206\\
		\hline		
		\end{tabular}
		\caption{\small{Isotopic composition of fuel bars for blue assembly in Fig.\ref{fig:carga_inicial_3colores}.  UO$_2$ and UO$_2$-Gd$_2$O$_3$ bars with an overall average enrichment of 1.76\% of $^{235}$U (relative to the total U). Fuel rods with Gd admixture have an enrichment of 5.2\% (in total) and 3.792\% (relative to U).}}
		\label{tab:ensamble_azul_tabla}
	\end{center}
\end{table}

The simulation was run simulating time intervals of 5 days of evolution of the nuclear reactor core until complete a full operation cycle of 400 days. 

As is expected for an initial load without plutonium, the fission rates of $^{239}$Pu and $^{241}$Pu start at zero and rapidly grow to become comparable to those of the U isotopes within a few weeks of operation. A more realistic situation should consider that between two fuel reloads, there will always be some remnant of plutonium inside the reactor, because when this action is performed, only a fraction of the spent fuel is replaced by new material.

\begin{table}[b]
	\begin{center} 
		\begin{tabular}{|l|r|} 
		\hline 
		Thermal Power (of each reactor) & 2027 MWth\\ 
		\hline
		Specific power (of each reactor) & 20.43 MW/Ton\\
		\hline
		Moderator & Demineralized water\\
		\ \ \  Temperature & 600 K\\
		\ \ \  Density & 0.720 g/cm$^3$\\
		\hline
		Coolant & Demineralized water\\
		\ \ \  Temperature & 400 K\\
		\ \ \  Density & 0.720 g/cm$^3$\\
		\hline
		Fuel assemblies  & 444\\ 
		\ \ \  Red assemblies & 68\\
		\ \ \  Blue assemblies & 96\\ 
	    \ \ \  Yellow assemblies & 280\\
		\hline
		Single unit cell (L$\times$L)& 1.295 cm$\times$1.295 cm\\ 
		\hline
		Fuel bars per assembly & 92\\
		\ \ \  Composition & UO$_2$/UO$_2$-Gd$_2$O$_3$\\
		\ \ \  Cladding & Zircaloy 2\\ 
	    \ \ \  Zircaloy density & 5.821 g/cm$^3$\\
		\ \ \  Zircaloy temperature & 600 K\\
		\ \ \  Fuel temperature & 900 K\\
		\ \ \  Fuel density  & 10.079 g/cm$^3$\\
		\ \ \  Fuel bar radius (R1)& 0.438 cm\\
		\ \ \  Zircaloy cladding radius (R2)& 0.513 cm\\
		\ \ \  Fuel bar height  & 400 cm\\
		\hline
		Coolant pipes per assembly & 2\\
		\ \ \  Composition & Zircaloy 2\\
		\ \ \  Pipe radius (occupied by water) & 1.0 cm\\
		\ \ \  Pipe radius (occupied by zircaloy) & 1.2 cm\\
		\hline
		\end{tabular}
		\caption{\small{Parameters of the BWR core used in the simulation with DRAGON code.}}
		\label{tab:parametros_simulacion}
	\end{center}
\end{table}

\section{\label{sec:Flujo}The antineutrino flux}

The power output of a nuclear reactor originates from the energy released in the fission of heavy elements, such as uranium and plutonium, into lighter fission fragments which are often unstable. The beta decay, $(A,Z)\rightarrow (A,Z+1)+e^-+\bar{\nu}_e$, of the fission fragments produces a large number of electron antineutrinos $\bar{\nu}_e$ which are emitted isotropically from the core.
During the fuel "burning" process, uranium isotopes breed $^{239}$Pu and $^{241}$Pu. The latter may in addition to the $^{235}$U undergo a fission process only with slow (thermal) neutrons, while $^{238}$U is fissionable by fast neutrons only. The decay of $^{239}$Pu produces substantially less antineutrinos than the decay of $^{235}$U in the same energy range, therefore during a typical reactor fuel cycle, the amount of antineutrinos decreases as uranium content decreases and the concentration of plutonium increases.

Each fission releases on average $\sim$200 MeV of energy and produces $\sim$6 $\bar{\nu}_e$ ($\sim 3$ beta decays per fission fragment), with energies below $\sim$10 MeV. This sets the number of antineutrinos emitted by a typical reactor to $\sim2\times10^{20} \bar{\nu}_e/s$ per GWth of thermal power. The energy spectrum of the antineutrinos depends on the fuel composition at a given time ($^{239}$Pu antineutrinos are slightly less energetic than those of $^{235}$U fission products).

The production of reactor antineutrinos is not exclusively through beta decay of the fission fragments of the four main fissile isotopes ($^{235}$U, $^{238}$U, $^{239}$Pu and $^{241}$Pu).The neutron capture in $^{238}$U(n,$\gamma$) $^{239}$U also generates this particles in a process that contributes about 17\% of the total antineutrino flux. This process occurs when a nucleus of $^{238}$U captures a neutron, leading to the following reaction: $^{238}$U+n$\rightarrow ^{239}$U $\rightarrow ^{239}$ Np $\rightarrow ^{239}$Pu, and produces two antineutrinos through two beta decays of $^{239}$U. This process occurs at a rate of $\sim$0.6 per fission, and produces antineutrinos with energies below 1.3 MeV.

There is no established experimental method for measuring the flux of very low energy antineutrinos. For a technological application such as reactor monitoring, only antineutrinos with energies above the threshold of the Inverse Beta Decay (IBD) reaction on free protons (E$_{\nu}$ \textgreater 1.8 MeV) can be realistically considered. IBD interactions of neutrinos on free protons can be identified through the correlation of the prompt positron signal and a delayed neutron capture signal. In general, the spectrum of the antineutrinos from a nuclear reactor detected via IBD has a peak between 3.5 and 4 MeV. Since the number of emitted antineutrinos and their average energy depend on the amounts of $^{235}$U, $^{238}$U, $^{239}$Pu and
$^{241}$Pu isotopes present in the core, the measured energy spectrum provides a direct image of the fuel composition of the reactor core at a given time.

Let us now consider the antineutrinos emitted by one of the reactors at the LVNPP. Ignoring the neutron capture contribution, the antineutrino flux $\Phi_i$ above 1.8 MeV can be calculated from the fission rates $f_i(t)$, and the antineutrino energy spectrum $S_i(E_{\bar{\nu}})$ of each isotope $i$, where $i = 1,2,3,4$ corresponds to each of the main fissile isotopes of the reactor core: $^{235}$U, $^{238}$U, $^{239}$Pu and $^{241}$Pu.

Assuming neutrinos are emitted isotropically, using a sphere of radius $R$, the flux at a time $t$ for each isotope $i$ is calculated as:

\begin{equation}
\Phi_i(E_{\bar{\nu}},t) = \frac{1}{4 \pi R^2} S_i(E_{\bar{\nu}}) f_i(t) ,
\label{eq:Flujoneutrinos}
\end{equation}
with $R$ the distance from the reactor core to the detector.

\begin{figure}[b] 
\centering    
\includegraphics[width=0.45\textwidth] {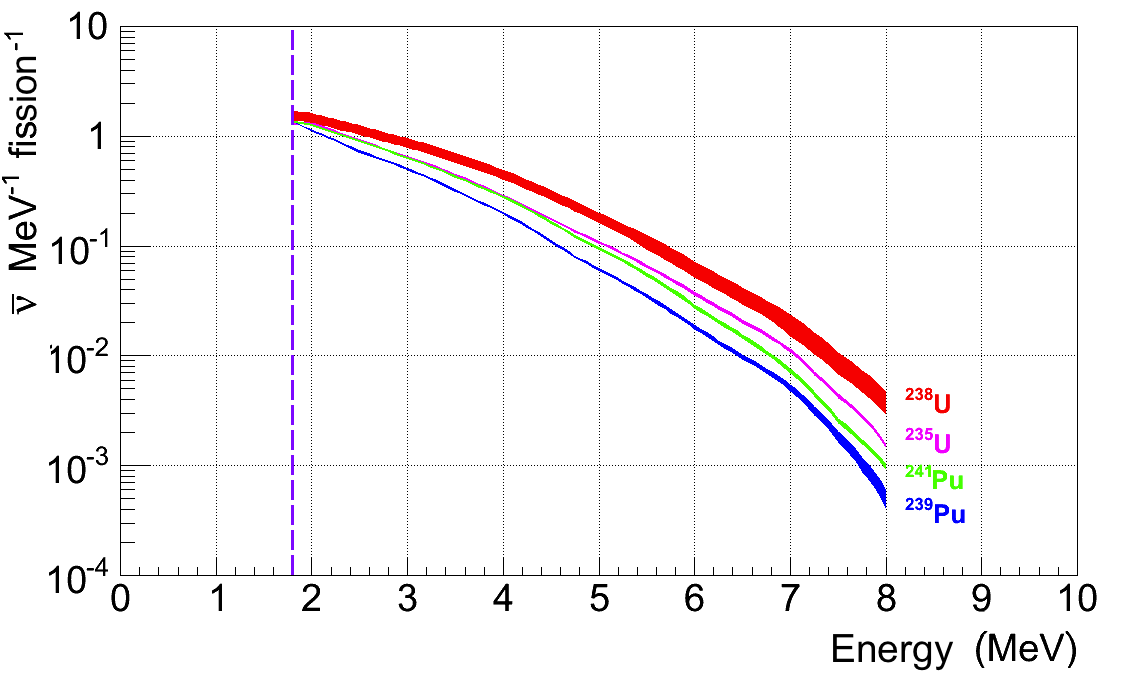}
\caption{\small{Energy spectra of the emitted antineutrinos per fission of $^{235}$U, $^{238}$U, $^{239}$Pu and $^{241}$Pu. The dotted line at 1.8 MeV indicates the threshold for the Inverse Beta Decay (IBD) process.}}
\label{fig:Mueller_Fallot}
\end{figure}

In our calculation of the antineutrino flux, we have used the new predictions of the energy spectra including the effect of the reactor antineutrino anomaly, as reported by~\cite{anomaly}, \cite{Fallot} and \cite{Mueller}. The antineutrino energy spectra $S_i(E_{\bar{\nu}})$ between 1.8 and 8 MeV are shown in Fig. \ref{fig:Mueller_Fallot}. For the purposes of reactor monitoring, the effect of this anomaly can be ignored as long as only comparisons between the relative changes of the spectrum throughout the reactor operation cycle are made.

\begin{figure}[t] 
\centering    
\includegraphics[width=0.5\textwidth] {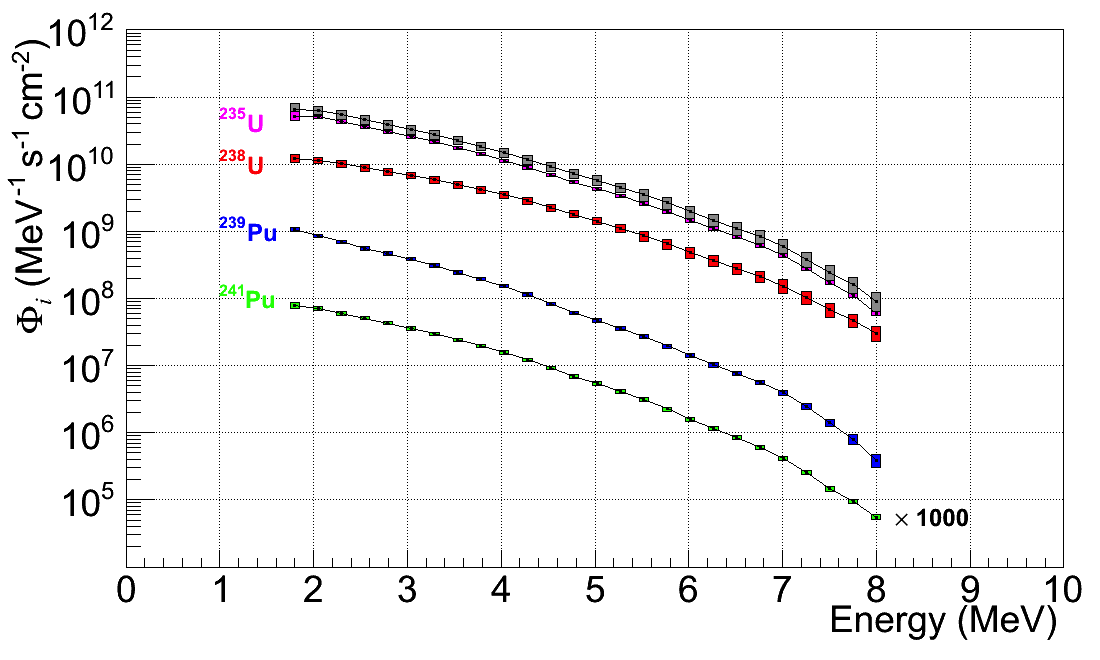}\\
\includegraphics[width=0.5\textwidth] {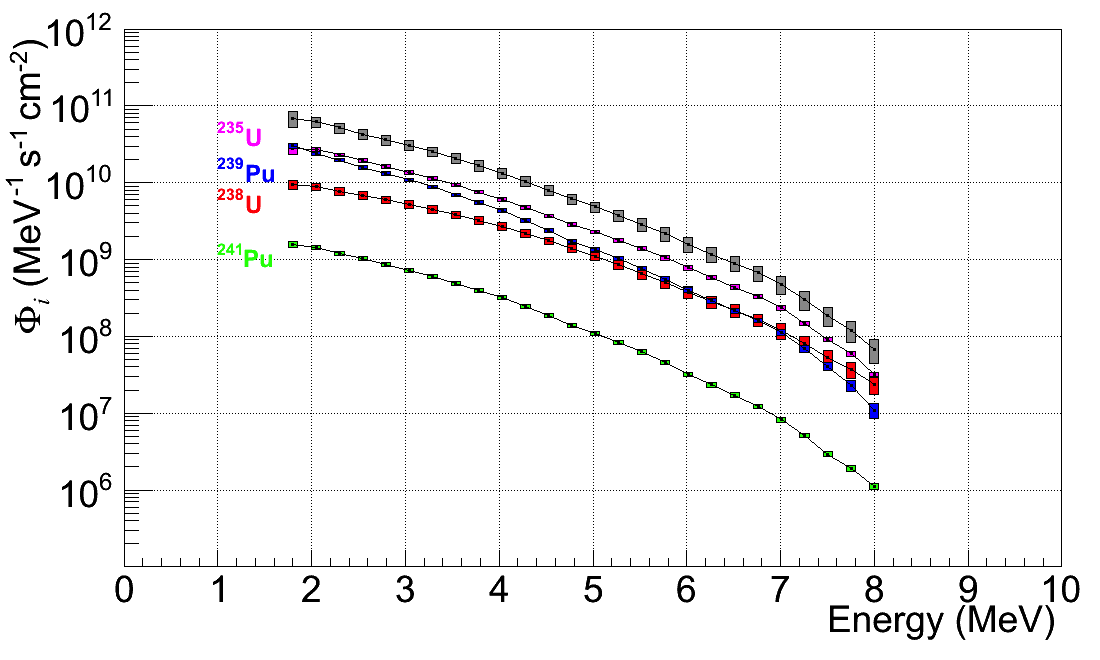}
\caption{\small{Antineutrino flux produced by the fission of the 4 main fissile isotopes at 100 meters from one of the LVNPP reactors of 2.027 GWth at day 5 (up) and 200 (bottom) of operation. The gray curve (on top) is the total flux.}}
\label{fig:flujos}
\end{figure}

Fig. \ref{fig:flujos} shows the antineutrino flux for each of the main fissile isotopes of one of the reactors in the LVNPP, at day 5 (top) and then at day 200 (bottom), in the latter the flux can be considered as stable. The total flux (the sum of the four isotopes) for each time period is also shown in the gray curve in this figure. The total flux at day 200 is shown in table \ref{tab:flujodia200}.

Two systematic uncertainties are associated with the calculation of the antineutrino flux: the error due to the energy spectra of antineutrinos $\delta(S(E_{\bar{\nu}}))$, and an error $\delta (P_{th})$ associated with fluctuations in the thermal power of the reactor core, assumed to be 5\% along one fuel cycle. The total error is the sum in quadrature of these uncertainties. The uncertainties in the antineutrino spectra are taken from Refs.~\cite{Fallot} and~\cite{Mueller}, and are of the order of 2-5\% for the dominant parts of the spectra from $^{235}$U,$^{239}$Pu, and $^{239}$Pu, and between 10-17\% for the dominant
part of the spectrum from $^{238}$U.

The total flux per reactor at day 200 from our calculation is compared in Fig. \ref{fig:flujoXsec200} to that obtained according to the parametrization suggested by Vogel and Engel~\cite{Xsec_DBI_vogel}, setting the fission rates of the four main fissile isotopes to these output by the simulation (2.7 $\times 10^{19}$ fis/s for $^{235}$U, 7.7 $\times 10^{18}$ fis/s for $^{238}$U, 2.8 $\times 10^{19}$ fis/s for $^{239}$Pu, and
1.5 $\times 10^{18}$ fis/s for 241 Pu). The two calculations agree well within the assumed uncertainties.

Spent fuel is stored temporarily in a nearby place to the reactor core and has a very small contribution to the antineutrino flux below the IBD reaction threshold, and was not considered for the calculation in this work.

\begin{table}[t]
	\begin{center} 
		\begin{tabular}{|l|c|c|} 
		\hline
		E [MeV] & $\Phi_{total}$ [MeV$^{-1}$ s$^{-1}$cm$^{-2}$]  & $\delta$(sist) [\%] \\ 
		\hline                    
		\hline
 1.800 & 6.921E+10 & 7.2\\ 
 2.048 & 6.192E+10 & 3.6\\ 
 2.296 & 5.185E+10 & 3.6\\
 2.544 & 4.304E+10 & 3.7\\
 2.792 & 3.634E+10 & 3.7\\
 3.040 & 3.052E+10 & 3.7\\ 
 3.288 & 2.542E+10 & 3.7\\ 
 3.536 & 2.066E+10 & 3.8\\ 
 3.784 & 1.671E+10 & 3.8\\
 4.032 & 1.336E+10 & 3.9\\ 
 4.280 & 1.045E+10 & 4.0\\ 
 4.528 & 8.026E+09 & 4.0\\ 
 4.776 & 6.147E+09 & 4.1\\ 
 5.024 & 4.841E+09 & 4.2\\ 
 5.272 & 3.761E+09 & 4.3\\ 
 5.520 & 2.871E+09 & 5.2\\ 
 5.768 & 2.157E+09 & 5.4\\
 6.016 & 1.584E+09 & 5.5\\
 6.264 & 1.179E+09 & 5.6\\
 6.512 & 8.900E+08 & 5.8\\ 
 6.760 & 6.692E+08 & 5.9\\ 
 7.008 & 4.752E+08 & 7.0\\ 
 7.256 & 3.027E+08 & 7.4\\ 
 7.504 & 1.881E+08 & 7.8\\ 
 7.752 & 1.210E+08 & 8.4\\
 8.000 & 6.763E+07 & 9.8\\
\hline
		\end{tabular}
		\
		\caption{\small{Total antineutrino flux at day 200 of one of the reactors of the LVNPP, at a location of 100 m from the reactor core.} }
		\label{tab:flujodia200}
	\end{center}
\end{table}

\section{\label{sec:eventos}Event Rates}

\begin{figure}[t] 
\centering    
\includegraphics[width=0.5\textwidth] {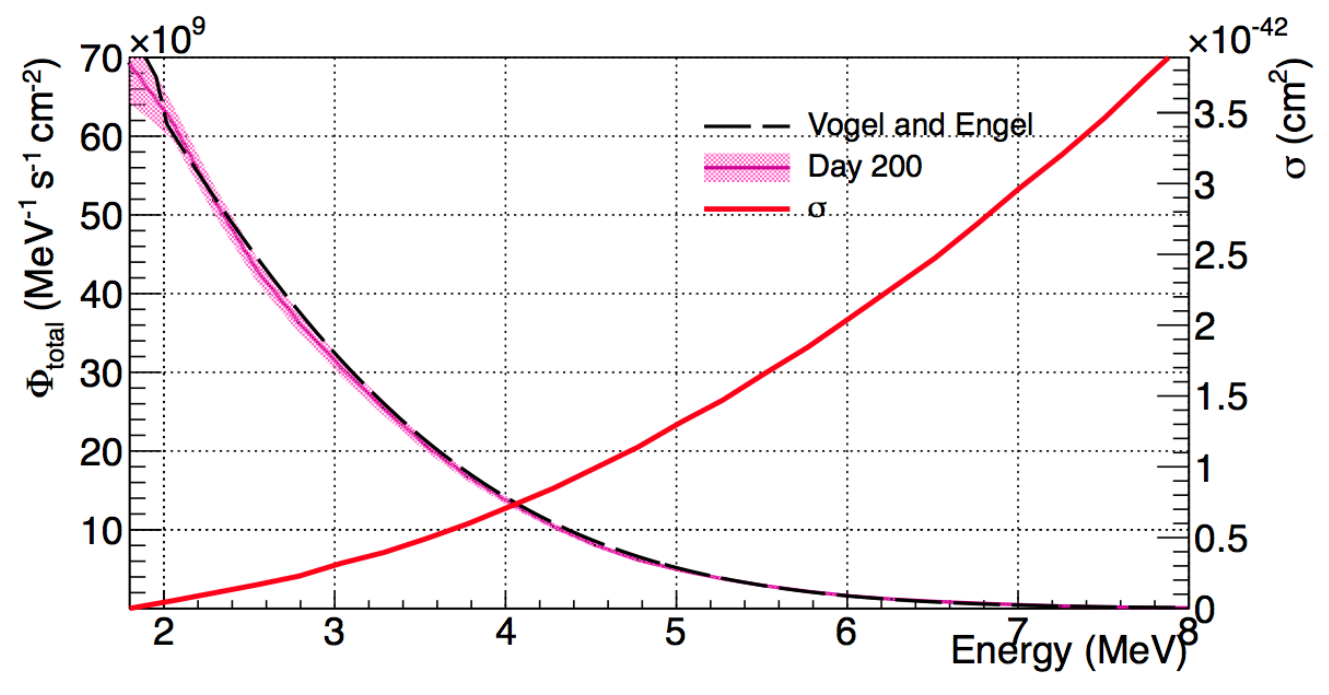}
\caption{\small{Total antineutrino flux per reactor at day 200 (solid curve with error band). The error includes the sum of the relative uncertainties of each of the 4 main fissile isotopes in the reactor. The dashed curve shows the parametrization suggested in~\cite{Xsec_DBI_vogel} with the fission rates calculated by DRAGON for the LVNPP. The continuous line shows the IBD cross section as given by~\cite{Angular_dist_Vogel}}}
\label{fig:flujoXsec200}
\end{figure}

Although the IBD reaction has a very small cross section ($\sim$10$^{-43}$cm$^2$), the enormous flux emitted by a nuclear reactor allows the signal to be observed with a relatively small detector located a few tens of meters from the reactor core. The total IBD cross-section as a function of the neutrino energy is shown in Fig. \ref{fig:flujoXsec200} (right scale). Here we consider 1 ton of polyvinyl-toluene (PVT) plastic scintillator located at 100 m from each of the reactor cores, similar to the detector considered in~\cite{PANDA}.

The number of IBD interactions associated with neutrinos from isotope $i$, emitted with energies between $E_{min}$ and $E_{max}$, and occurring in a time interval $\Delta t$ ($\Delta N_{\textrm{ev} (i)}$) is calculated by integrating the product of the flux $\Phi_i$, the IBD cross section $\sigma(E_{\bar{\nu}})$ \cite{Angular_dist_Vogel}, and the number of targets (in this case free protons) $N_p$, over the energy interval:

\begin{equation}
\frac{\Delta N_{\textrm{ev} (i)}}{\Delta t}= N_{p} \displaystyle\int_{E_{min}}^{E_{max}} \Phi_i(E_{\bar{\nu}},t) \sigma (E_{\bar{\nu}}) \, dE_{\bar{\nu}} .
\label{eq:Neventosi}
\end{equation}

Depending on the characteristics of the detector, only a fraction of these interactions will be observed. For a simple estimate, let this fraction be a uniform efficiency factor $\epsilon$. For neutrinos coming from nuclear reactors, $E_{min}$ is the threshold of 1.8 MeV and $E_{max}$ is $\sim~$10~MeV. The number of detected events for each isotope is then approximated by:

\begin{equation}
N_{\textrm{ev}(i)} =\epsilon \times N_{p} \frac{\Delta t}{4 \pi R^2}\displaystyle\int_{E_{min}}^{E_{max}}  S_i(E_{\bar{\nu}}) f_i(t) \sigma (E_{\bar{\nu}}) \, dE_{\bar{\nu}} ,
\end{equation}

where $\Delta t$ is a small exposure time compared with the time scale in which fission rates $f_i (t)$ change appreciably. Calculation for longer times requires an integration over time.

Finally the total number of events is equal to the sum of the events for each isotope:

\begin{equation}
N_{\textrm{ev (total)}} = \displaystyle\sum_{i=1}^4 N_{\textrm{ev} (i)} 
\end{equation}

\begin{figure}[t] 
\centering    
\includegraphics[width=0.5\textwidth] {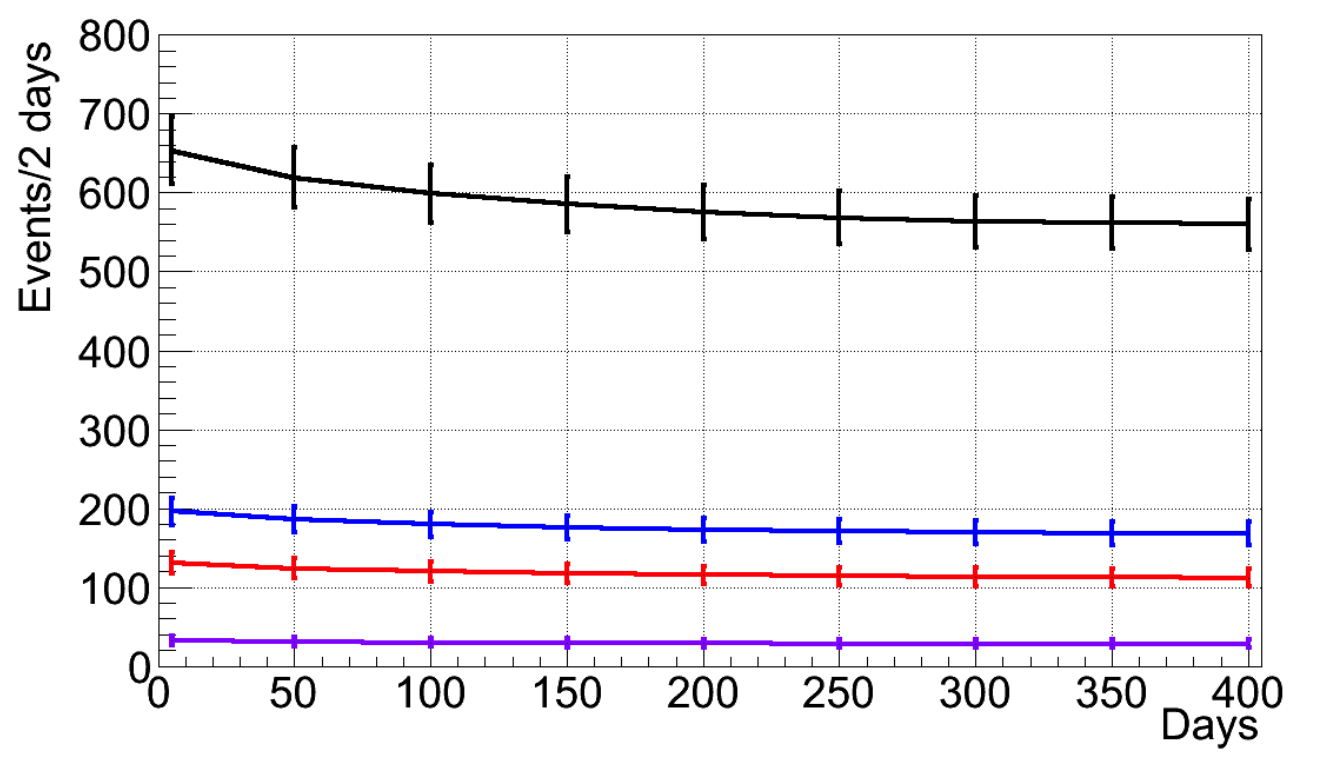}
\caption{\small{Number of events accumulated over 2 days of exposure along a complete cycle of 400 days at LVNPP with two reactors in 1 ton of plastic scintillator. Curves for detection efficiencies (lower to higher) of 5\%, 20\%, 30\%, and 100\% are shown}}
\label{fig:eventostodos}

\end{figure}

These are shown in Fig. \ref{fig:eventostodos}, considering three different detector efficiencies (5\%, 20\%, 30\% and 100\%) and counting for two consecutive days.

The authors of~\cite{PANDA} estimate an antineutrino detection efficiency of 9.24\% and an antineutrino event rate of 147 events/day in a 1 ton detector with an uncertainty of the order of 30\%. This is consistent, within errors, with our expected rate of 280 events/day (stable after day 350), once scaling for reactor power, distance, and assuming the same efficiency. The authors measured a background of $\sim$365 events/day, operating a 360 kg detector with the reactor OFF for 21 days. Scaling to a 1 ton detector corresponds to $\sim$1013 events/day. Since our expected rate at LVNPP is 26 events/day at the location of 100 m from the reactor cores, a background measurement at the level of 0.5\% ($\pm$5 events/day) would be necessary for a 5$\sigma$ detection of the antineutrinos, comparing the ON/OFF event rates with a similar detector.

\section{\label{sec:Conclusiones}Conclusions}

We calculated the antineutrino flux produced by the reactors at the LVNPP in Mexico using the DRAGON simulation code in the energy range 1.8 MeV\textless E$_{\nu}$\textless 8 MeV. The simulation provided the time evolution of the fission rates and hence the change of the flux along the 400 day long fuel cycle, starting with an initial load with no plutonium. A total systematic uncertainty on the flux ranging from 3.6\% to 9.8\%, depending on the antineutrino energy, was calculated assuming a 5\% uncertainty on the reactor power, and the uncertainties reported for the antineutrino spectra of the dominant fissile isotopes (~\cite{Fallot},\cite{Mueller}).Our calculation agrees well with a frequently parametrization \cite{Xsec_DBI_vogel} when the fission rates output by the simulation of the LVNPP BWR reactors are used.

The simulation code required the knowledge of generic parameters of the BWR reactors, at LVNPP, available in the literature, and showed the fission rates to have a moderate dependence on variations of the coolant, moderator and fuel temperatures for a fixed thermal power.

A segmented plastic scintillator detector of the type considered in~\cite{PANDA} located at 100 m from the two reactors could detect the neutrinos emitted, provided a background measurement with the reactors OFF is performed at the 0.5\% level.

\section*{\label{sec:Agradecimientos}Acknowledgements}
The authors acknowledge the support of UNAM-PAPIIT, grants number IB100413, IN112213 and CONACYT through grants CB-2009/00131598 and 398495 (scholarship holder number: 570620).
We also especially thank Dr. Guy Marleau from \'Ecole Polytechnique de Montr\'eal and Dr. Christopher Jones from MIT for their valuable help and advice throughout the installation and implementation of the DRAGON code simulation.


\begin{thebibliography}{99}

%\bibitem {Cowan_Reines} F. Reines, C. L. Cowan, \textit{Detection of the free neutrino} Phys. Rev. vol 92, 830 (1953)
\bibitem {DayaBay} The Daya Bay Collaboration \textit{Observation of electron-antineutrino disappearance at Daya Bay} Physical Review Letters 108, 17 (2012)
\bibitem {DoubleCHOOZ} Y. Abe et al. (Double Chooz Collaboration) \textit{Indication of Reactor $\bar{\nu}_e$ Disappearance in the Double Chooz Experiment} Phys. Rev. Lett. 108, 131801 (2012)  
\bibitem{SONGS} N.S. Bowden, A.Bernstein, S. Dazeley, R.Svoboda, A. Misner and T.Palmer \textit{Observation of the Isotopic Evolution of Pressurized Water Reactor Fuel Using an Antineutrino Detector} J.Appl.Phys. 105  064902 
\bibitem {PANDA} Y. Kuroda, S. Oguri et al., \textit{A mobile antineutrino detector with plastic scintillators}, Nuclear Instruments and Methods in Physics Research Section A: Accelerators, Spectrometers, Detectors and Associated Equipment, Vol 690,41-47,(2012). S. Orugi, Y. Kuroda, Y. Kato, R. Nakata, Y. Inoue, C. ito, M. Minowa, \textit{Reactor antineutrino monitoring with a plastic scintillator array as a new safeguards method},  Nuclear Instruments and Methods in Physics Research Section A: Accelerators, Spectrometers, Detectors and Associated Equipment, Vo. 757, 33-39, (2014).
\bibitem {Nucifer} A Porta (for the Nucifer collaboration)  \textit{Reactor neutrino detection for non proliferation with the Nucifer experiment} J. Phys.: Conf. Ser. 203 012092 (2010)
\bibitem {Huber} Christensen E., Huber P., Jaffke P., \textit{Antineutrino Monitoring for Heavy Water Reactors}, Phys. Rev. Lett. 113, 042503 (2014).
\bibitem {ANGRA} E Casimiro and J C Anjos, \textit{Cosmic muon background and reactor neutrino detectors: the Angra experiment}  J. Phys.: Conf. Ser. 116 012003 (2008)
\bibitem {DANSS} V.Belov et al, \textit{Registration of reactor neutrinos with the highly
segmented plastic scintillator detector DANSSino} Journal of Instrumentation, 8 P05018 (2013)

\bibitem {IAEA_workshop} \textit{IAEA Headquarters. Final Report: Focused Workshop on
Antineutrino Detection for safeguards Applications.} October 2008.

\bibitem {literature} C. O. Muehlhause and S. Oleksa., \textit{Antineutrino Flux from a Reactor}, Phys. Rev. 105, 1332 (1957). A.I Afonin, S.A.Bogatov et al.,\textit{Search for neutrino oscillations in an experiment in the reactor of the Rovno nuclear power plant.} Pis’ma Zh. Eksp. Teor. Fiz.42, No. 5, 230-233 (1985).G. Zacek, F. v. Feilitzsch et al., \textit{Neutrino-oscillation experiments at the Gsgen nuclear power reactor.} Phys.Rev. D 34, 2621 (1986). Y. Declaisa, H. de Kerretb, et al. \textit{Study of reactor antineutrino interaction with proton at Bugey nuclear power plant.} Phys. Lett. B. 338, 2-3,27, 383-389 (1994). V.I. Kopeikin, L.A. Mikaelyan, V.V. Sinev. \textit{Spectrum of electronic reactor antineutrinos.} Phys.Atom.Nucl. 60, 172-176 (1997). M. Apollonio et al.\textit{Search for neutrino oscillations on a long baseline at the CHOOZ nuclear power station.} Eur.Phys.J. C27, 331-374 (2003). G. Mention, M. Fechner et al. \textit{Reactor antineutrino anomaly.} Phys. Rev. D 83, 073006 (2011). Jun Cao \textit{Determining Reactor Neutrino Flux.} Nuclear Physics B Proceedings Supplement 00 (2012).


\bibitem {Nakajima} K. Nakajima, et al., \textit{A simple model of reactor cores for reactor neutrino flux calculations for the KamLAND experiment.} Nuclear Instruments and Methods in Physics Research A 569 (2006) 837-844.


\bibitem{reporteIAEA} \textit{Energy, Electricity and Nuclear Power Estimates for the Period up to 2050}. Reference data series No. 1, IAEA, 2014 Edition
\bibitem{anomaly} G. Mention et al., \textit{The Reactor Antineutrino Anomaly}, Phys. Rev. D83, 073006 (2011).
\bibitem {Fallot} M. Fallot et al., \textit{New antineutrino energy spectra predictions from the summation of beta decay branches of the fission products}, Phys. Rev. Lett. 109, 202504 (2012).
\bibitem {Mueller} Th. A. Mueller et al., \textit{Improved predictions of reactor antineutrino spectra}, Phys. Rev. C 83, 054615 (2011).
\bibitem {Pagina_DRAGON} G. Marleau, R. Roy and A. H\'ebert, \textit{DRAGON: A Collision Probability Transport Code for Cell and Supercell Calculations}, Report IGE-157, Institut de g\'enie nucl\'eaire, \'Ecole Polytechnique de Montr\'eal, Montr\'eal, Qu\'ebec (1994); DRAGON code website. http://www.polymtl.ca/nucleaire/DRAGON/en/
\bibitem {modifica_DRAGON} Modifications to the installation of DRAGON http://dspace.mit.edu/handle/1721.1/70045

\bibitem {fiss_frac_DRAGON} X.B. Ma, L.Z. Wang, Y.X. Chen, W.L. Zhong, F.P. An, \textit{Uncertainties analysis of fission fraction for reactor antineutrino experiments using DRAGON}, arXiv:1405.6807 (2014)


\bibitem {folletoCNLV} \textit{\textquestiondown Qu\'e es el ciclo del combustible nuclear?} Booklet edited by CFE. F-65.
\bibitem {CJones-DRAGON_MURE} C. L. Jones, A. Bernstein et al., \textit{Reactor simulation for antineutrino experiments using DRAGON and MURE}, Phys. Rev. D 86, 012001 (2012).
\bibitem {Hernandez} H. H. L\'opez,   \textit{Fuel asembly with inert matrix fuel rods as reload options for Laguna Verde NPP}, Annals of Nuclear Energy, vol. 40, 215-220 (2012).
\bibitem {Xsec_DBI_vogel} P.Vogel, J. Engel, Phys. Rev. D., 39, 3378 (1989).
\bibitem {Angular_dist_Vogel} P. Vogel, J. F. Beacom, \textit{Angular distribution of neutron
inverse beta decay, $\nu_e + p \rightarrow e^+ + n$}, Phys. Rev. D, vol 60, 053003 (1999).
  

\end{thebibliography}
\end{document}